\documentclass[prd,aps,preprintnumbers,nofootinbib,superscriptaddress]{revtex4}
\pdfoutput=1
\usepackage{amssymb,amsmath,graphicx,bm}

\usepackage[usenames,dvipsnames]{color}
\usepackage[normalem]{ulem} 
\renewcommand\sout{\bgroup \color[rgb]{0.55,0.00,0.99} \ULdepth=-.5ex \ULset}
\newcommand{\xB}{x_{\scriptscriptstyle B}}

\newcommand{\sT}{{\scriptscriptstyle T}}
\renewcommand{\d}{\mathrm{d}}
\def\slash#1{\setbox0=\hbox{$#1$}               
        \dimen0=\wd0                            
        \setbox1=\hbox{/} \dimen1=\wd1          
        \ifdim\dimen0>\dimen1                   
        \rlap{\hbox to \dimen0{\hfil/\hfil}}    
        #1                                      
        \else
        \rlap{\hbox to \dimen1{\hfil$#1$\hfil}} 
        /                                       
        \fi}                                    %

\usepackage[normalem]{ulem} 

\begin{document}

\title{Extracting color octet NRQCD matrix elements from $J/\psi$ production at the EIC}

\author{Dani\"el Boer}
\email{d.boer@rug.nl}
\affiliation{Van Swinderen Institute for Particle Physics and Gravity, University of Groningen, Nijenborgh 4, 9747 AG Groningen, The Netherlands}

\author{Cristian Pisano}
\email{cristian.pisano@unica.it}
\affiliation{Dipartimento di Fisica, Universit\`a di Cagliari, Cittadella Universitaria, I-09042 Monserrato (CA), Italy}
\affiliation{INFN Sezione di Cagliari,  Cittadella Universitaria, I-09042 Monserrato (CA), Italy}

\author{Pieter Taels}
\email{pieter.taels@polytechnique.edu}
\affiliation{Centre de Physique Th\'eorique, \'Ecole polytechnique, CNRS, I.P.~Paris, F-91128 Palaiseau, France}

\begin{abstract}
Recently unpolarized and polarized $J/\psi \,(\Upsilon)$ production at the Electron-Ion Collider (EIC) has been proposed as a new way to extract two poorly known color-octet NRQCD long-distance matrix elements: $\langle0\vert{\cal O}_{8}^{J/\psi}(^{1}S_{0})\vert0\rangle$ and $\langle0\vert{\cal O}_{8}^{J/\psi}(^{3}P_{0})\vert0\rangle$. The proposed method is based on a comparison to open heavy-quark pair production ideally performed at the same kinematics. In this paper we analyze this proposal in more detail and provide predictions for the EIC based on the available determinations of the color-octet matrix elements. We also propose two additional methods that do not require comparison to open heavy-quark pair production.
\end{abstract}

\date{\today}

\maketitle

\section{Introduction and formalism}
Semi-inclusive $J/\psi$ and $\Upsilon$ production in deep-inelastic lepton-proton scattering, i.e.\ $e\,p \to e^\prime \,J/\psi \,(\Upsilon)\,X$, where both the electron and the proton are unpolarized, is sensitive to unpolarized as well as linearly polarized transverse momentum dependent gluon distributions (TMDs)~\cite{Mulders:2000sh,Meissner:2007rx,Boer:2016xqr}. The latter lead to azimuthal $\cos 2\phi_{\scriptscriptstyle{T}}$ modulations of the cross section differential in the transverse momentum of the $J/\psi$ or $\Upsilon$ w.r.t.\ the lepton scattering plane \cite{Mukherjee:2016qxa,Rajesh:2018qks,Bacchetta:2018ivt}.
Unfortunately, predictions for the asymmetries are hampered by the dependence on two poorly known color-octet (CO) NRQCD long-distance matrix elements: $\langle0\vert{\cal O}_{8}^{J/\psi}(^{1}S_{0})\vert0\rangle$ and $\langle0\vert{\cal O}_{8}^{J/\psi}(^{3}P_{0})\vert0\rangle$. 
In Ref.~\cite{Bacchetta:2018ivt}, taking ratios with spin-dependent asymmetries was suggested to cancel out these quantities. 
It was also pointed out in Ref.~\cite{Bacchetta:2018ivt} that ratios with the analogous expressions for unpolarized open heavy-quark pair production could be used to cancel out the gluon TMDs so as to determine the CO matrix elements experimentally. This can hopefully help reducing their uncertainty and, as a consequence, forgo the need for comparison to spin asymmetries. In this paper we look into the determination of CO matrix elements in more detail, providing the general expressions and their current estimates, which can be of help in the study of these quantities at the EIC.  

Following the calculations of Ref.~\cite{Bacchetta:2018ivt}, we study the semi-inclusive  deep-inelastic (SIDIS) process of quarkonium production
\begin{equation}
e(\ell) + p(P) \to e(\ell^{\prime}) + {\cal Q}\,(P_{\cal Q}) + X\,,
\end{equation}
where ${\cal Q}$ is either a $J/\psi$ or a $\Upsilon$ meson. Besides unpolarized quarkonium production we also examine the cases in which the quarkonium is longitudinally or transversely polarized w.r.t.\ its direction of motion in the $\gamma^*p$ center-of-mass frame. The  reference frame is such that both the exchanged virtual photon $\gamma^*$ and the incoming proton move along the $\hat z$-axis. Azimuthal angles are measured w.r.t.\ to the lepton scattering plane. In order for TMD factorization to apply, the component of the quarkonium momentum transverse to the lepton plane, i.e.\ $q_\sT \equiv  P_{{\cal Q} \sT}$, should be small compared to the virtuality of the photon $Q$ and to the mass of the quarkonium $M_{\cal Q}$\footnote{For recent studies of the large $p_\sT$-range probed in $J/\psi + {\rm jet}$ production, see Refs.~\cite{DAlesio:2019qpk,Kishore:2019fzb}.} The differential cross section can then be written as 
\begin{eqnarray}
\d\sigma^{[e \, p\, \to \, e^\prime\, {\cal Q}_{P}\, X]}
& = &\frac{1}{2 s}\,\frac{\d^3 \ell'}{(2\pi)^3\,2 E_e^{\prime}} \frac{\d^3 P_{\cal Q}}{(2\pi)^3\,2 E_{\cal Q}}
{\int}\d x\, \d^2\bm p_{\sT}\,(2\pi)^4
\delta^4(q {+} p {-} P_{\cal Q})
 \nonumber \\
&&\qquad \qquad\qquad \qquad\qquad\times  \frac{1}{x^2\,Q^4}\, 
L_{\mu \rho}(\ell,q) \,  \Gamma_{g\,\nu\sigma}(x{,} \bm p_{\sT})
\, H^{\mu\nu}_{\gamma^*\, g \rightarrow {\cal Q}_{P} } \,H^{\star\, \rho\sigma}_{\gamma^*\, g \rightarrow {\cal Q}_{P} } \, , 
\label{CrossSec}
\end{eqnarray}
where $ \Gamma_g$ is the gluon correlator and $L(\ell, q)$ is the lepton tensor. For details of the intermediate calculation we refer to  Ref.~\cite{Bacchetta:2018ivt}. Sticking to the notation of Ref.~\cite{Bacchetta:2018ivt}, the differential cross section will be denoted by 
\begin{equation}
\d\sigma^{UP} (\phi_\sT) \equiv \frac{\d\sigma^{[e \, p\, \to \, e^\prime\, {\cal Q}_{P}\, X]}}
{\d z\,\d y\,\d\xB\,\d^2\bm{q}_{\sT}}  \,,
\label{eq:cs}
\end{equation}
which depends on the variable $z = P\cdot P_\psi/P\cdot q$, on the inelasticity $y = P\cdot q/P\cdot \ell$, on Bjorken-$x$  $\xB = Q^2/2P\cdot q$, and on the transverse momentum $\bm{q}_{\sT} = \bm P_{{\cal Q}\sT}$ of the quarkonium state ${\cal Q}$ ($P_{{\cal Q}\sT}^2 = -\bm P_{{\cal Q}\sT}^2$) with an azimuthal angle $\phi_\sT$ w.r.t.\ the lepton plane ($\phi_{\ell}=\phi_{\ell^\prime}=0$). The variables $y$ and $\xB$ are related to
the total invariant mass squared $s =  (\ell + P)^2 \approx 2\,\ell\cdot P$ and to the photon virtuality $Q^2 = -q^2 \equiv -(\ell-\ell^\prime)^2$ through the relation 
$s = 2\,P\cdot q/y = Q^2/\xB y$.
The superscript $UP=UU, UL $ or $UT$ denotes an unpolarized proton and a polarization state $P$ for the quarkonium which can be either unpolarized ($U$), longitudinally polarized ($L$) or transversely polarized  ($T$) with respect to the direction of its three-momentum in the photon-proton center-of-mass frame.

For the calculation of the cross section, we adopt the TMD factorization framework in combination with nonrelativistic QCD (NRQCD)~\cite{Hagler:2000dd,Yuan:2000qe,Yuan:2008vn}. In NRQCD, the nonperturbative hadronization process of the heavy-quark pair into a quarkonium bound state is encoded in long-distance matrix elements (LDMEs)~\cite{Bodwin:1994jh}. Observables are evaluated by means of a double expansion in the strong coupling constant $\alpha_s$ and in the average velocity $v$ of the heavy quark in the quarkonium rest frame~\cite{Lepage:1992tx}, where $v^2 \simeq 0.3$ for charmonium and $v^2 \simeq 0.1$ for bottomonium. In SIDIS to leading order (LO) in  $\alpha_s$, the heavy quark-antiquark pair can only be produced in a color-octet (CO) state. The CO LDMEs are typically determined from fits to data on $J/\psi$ and $\Upsilon$ yields~\cite{Butenschoen:2010rq,Chao:2012iv,Sharma:2012dy,Bodwin:2014gia,Zhang:2014ybe}. The extracted values are not compatible with each other, even within the large uncertainties. Therefore, any new method to determine the CO matrix elements with better precision is worth exploring. Moreover, the methods to be discussed partly rely on experimental determination of the polarization states of the produced quarkonia, which is not only beneficial for the extraction of the CO matrix elements but also to the understanding of the production mechanism, which is still incomplete. Although NRQCD successfully explains many experimental observations, describing all cross sections and polarization measurements for charmonia in a consistent way still poses challenges~\cite{Brambilla:2010cs,Andronic:2015wma,Lansberg:2019adr}.

In the process under study, at LO in the strong coupling constant $\alpha_s$,  the $Q \overline Q$ pair forms a bound state with  spin $S$, orbital angular momentum $L$, and total angular momentum $J$, for which we employ spectroscopic notation: $^{2S+1}L_J$. Since to LO accuracy there is no gluon emission in the final state, $z=1$. The relevant CO matrix elements for $J/\psi$ production are  $\langle 0 \vert {\cal O}_8^{J/\psi} (^1 S_0)\vert 0 \rangle$ and $\langle 0 \vert {\cal O}_8^{J/\psi} (^3 P_J)\vert 0 \rangle$, with $J=0,1,2$, and the subscript $8$ denotes the color configuration. The color singlet (CS) production mechanism is possible only at ${\cal O}(\alpha_s^2)$, where the $Q \overline Q$ is formed in a $^3S_1^{(1)}$ state together with a gluon, hence $0 \le z \le 1$. The CS contribution is, therefore, suppressed relatively to the CO one by a perturbative coefficient of the order $\alpha_s/\pi$~\cite{Fleming:1997fq,Qiu:2020xum} and vanishes at $z=1$. On the other hand, $\langle 0 \vert {\cal O}_8^{J/\psi} (^1 S_0)\vert 0 \rangle$ and $\langle 0 \vert {\cal O}_8^{J/\psi} (^3 P_J)\vert 0 \rangle$ are suppressed compared to $\langle 0 \vert {\cal O}_1^{J/\psi} (^3 S_1)\vert 0 \rangle$ by $v^3$ and $v^4$,  respectively.  As a result the CO contribution should be enhanced by about a factor $v^3 \pi/\alpha_s \approx 2$ w.r.t.\ the CS contribution. In the analysis of Ref.~\cite{Fleming:1997fq} this factor turned out to be larger ($\approx 4$) for values of  $Q^2 > 4$ GeV$^2$. Based on these findings, we restrict our study to the CO contributions at leading order in $\alpha_s$.

The cross section for $e \, p\, \to \, e^\prime\, {\cal Q}_{P}\, X$ in Eq.~\eqref{eq:cs} can be cast in the following form:
\begin{align}
\d\sigma^{UP}
  & =  {{\cal N}}\, \bigg [ A^{UP}  f_1^g (x, \bm q_\sT^2 )+  \frac{\bm q_\sT^2}{M_p^2}\, B^{UP}\, h_1^{\perp\, g} (x, \bm q_\sT^2 ) \cos 2 \phi_\sT  \bigg ] \delta(1-z)\,,
\label{eq:csU}
\end{align}
where the unpolarized gluon TMD $f_1^g$ and the linearly polarized gluon TMD $h_1^{\perp\, g}$ depend, besides on the transverse momentum, also on the momentum fraction $x$ given by
\begin{align}
 x & = \xB + \frac{M^2_{\cal Q}}{y\,s}  = \frac{M^2_{\cal Q} + Q^2}{y\,s} =    \xB\,\frac{M^2_{\cal Q} + Q^2}{Q^2} \, .
 \label{eq:xresult}
\end{align}
The normalization factor $\cal N$ reads
\begin{equation}
{\cal N} =   (2 \pi)^2  \frac{{\alpha^2 \alpha_se_Q^2}}{ y\,  Q^2\, M_{\cal Q}(M_{\cal Q}^2+Q^2)} \,,
\label{eq:N}
\end{equation}
with $e_Q$ denoting the fractional electric charge of the quark $Q$.
The expressions for $A^{UP}$ and $B^{UP}$ are
\begin{align}
{A}^{UP} 
  = &~  [1+(1-y)^2]\,{\cal A}_{U+L}^{\gamma^*g\to {\cal Q}_P}  \, - \,   y^2\, {\cal A}_{L}^{\gamma^*g \to {\cal Q}_P} \,,
\label{eq:AgstarP}  \\
{B}^{UP} =
 &  ~  (1-y)\, {\cal B}_{T}^{\gamma^*g \to {\cal Q}_P}\,.
\end{align} 
where the subscripts $U+L$, $L$, $T$ refer to the specific polarization of the photon~\cite{Pisano:2013cya,Brodkorb:1994de}. 
Using heavy-quark spin symmetry relations~\cite{Bodwin:1994jh}  
\begin{equation}
 \langle 0 \vert {\cal O}_8^{J/\psi} (^3 P_J)\vert 0 \rangle=(2J+1)\langle 0 \vert {\cal O}_8^{J/\psi} (^3 P_0)\vert 0 \rangle\, + \,{{\cal O}(v^2)}\,,
 \label{eq:hqss}
 \end{equation}
one obtains to leading order in $v$:
\begin{align}
{\cal A}_{U+L}^{\gamma^{*}g\to{\cal Q}_{U}}= & \langle0\vert{\cal O}_{8}^{J/\psi}(^{1}S_{0})\vert0\rangle+\frac{12}{N_{c}}\frac{7M_{{\cal Q}}^{2}+3Q^{2}}{M_{{\cal Q}}^{2}(M_{{\cal Q}}^{2}+Q^{2})}\langle0\vert{\cal O}_{8}^{J/\psi}(^{3}P_{0})\vert0\rangle\,,\\
{\cal A}_{L}^{\gamma^{*}g\to{\cal Q}_{U}}= & \frac{96}{N_{c}}\,\frac{Q^{2}}{(M_{{\cal Q}}^{2}+Q^{2})^{2}}\langle0\vert{\cal O}_{8}^{J/\psi}(^{3}P_{0})\vert0\rangle\,,\\
{\cal B}_{T}^{\gamma^{*}g\to{\cal Q}_{U}}= & -\langle0\vert{\cal O}_{8}^{J/\psi}(^{1}S_{0})\vert0\rangle+\frac{12}{N_{c}}\frac{3 M_{{\cal Q}}^{2}-Q^{2}}{M_{{\cal Q}}^{2}(M_{{\cal Q}}^{2}+Q^{2})}\, \langle0\vert{\cal O}_{8}^{J/\psi}(^{3}P_{0})\vert0\rangle\, .
\end{align}
Similarly, the expressions for longitudinally polarized quarkonium production are
\begin{align}
{\cal A}_{U+L}^{\gamma^{*}g\to{\cal Q}_L}= &\, \frac{1}{3}\, \langle0\vert{\cal O}_{8}^{J/\psi}(^{1}S_{0})\vert0\rangle+\frac{12}{N_{c}}\frac{M_{{\cal Q}}^{4}+ 10 \, M_{\cal Q}^2\,  Q^2+Q^{4}}{M_{{\cal Q}}^{2}(M_{{\cal Q}}^{2}+Q^{2})^2}\langle0\vert{\cal O}_{8}^{J/\psi}(^{3}P_{0})\vert0\rangle\,,
\label{eq:AUL_L}\\
{\cal A}_{L}^{\gamma^{*}g\to{\cal Q}_L}=  &\,   {\cal A}_{L}^{\gamma^{*}g\to{\cal Q}} = \frac{96}{N_{c}}\,\frac{Q^{2}}{(M_{{\cal Q}}^{2}+Q^{2})^{2}}\langle0\vert{\cal O}_{8}^{J/\psi}(^{3}P_{0})\vert0\rangle\,, \\
{\cal B}_{T}^{\gamma^{*}g\to{\cal Q}_L}= & \, -\frac{1}{3}\, \langle0\vert{\cal O}_{8}^{J/\psi}(^{1}S_{0})\vert0\rangle+\frac{12}{N_{c}}\frac{1}{M_{{\cal Q}}^{2}} \, \langle0\vert{\cal O}_{8}^{J/\psi}(^{3}P_{0})\vert0\rangle\, .
\end{align}
Since $\d\sigma^{UU} = \d\sigma^{UL}+\d\sigma^{UT}$, this determines the corresponding expressions for transverse polarization of the quarkonium:
\begin{align}
{\cal A}_{U+L}^{\gamma^{*}g\to{\cal Q}_T}= &\, \frac{2}{3}\, \langle0\vert{\cal O}_{8}^{J/\psi}(^{1}S_{0})\vert0\rangle+\frac{24}{N_{c}}\frac{3\, M_{{\cal Q}}^{4}+Q^{4}}{M_{{\cal Q}}^{2}(M_{{\cal Q}}^{2}+Q^{2})^2}\langle0\vert{\cal O}_{8}^{J/\psi}(^{3}P_{0})\vert0\rangle\,,\\
{\cal A}_{L}^{\gamma^{*}g\to{\cal Q}_T}=  &\,  0\, ,\\
{\cal B}_{T}^{\gamma^{*}g\to{\cal Q}_T}= & \, -\frac{2}{3}\, \langle0\vert{\cal O}_{8}^{J/\psi}(^{1}S_{0})\vert0\rangle+ \frac{24}{N_{c}}\frac{1}{M_{{\cal Q}}^{2}} \,  \frac{M_{\cal Q}^2-Q^2}{M_{\cal Q}^2+Q^2} \, \langle0\vert{\cal O}_{8}^{J/\psi}(^{3}P_{0})\vert0\rangle\, .
\end{align}

One can isolate the angular independent term of the cross section and the azimuthal modulation by (weighted) integration:  
\begin{align}
D^{{\cal Q}_{P}}  & \equiv   \int \d \phi_\sT \,\frac{\d\sigma^{UP}}{\d y\,\d\xB\,\d^2 \bm{q}_{\sT} } =   2 \pi\, {\cal N}\, A^{UP}\,f_1^g(x, \bm q_\sT^2)\,, \label{eq:D-Na}\\
N^{{\cal Q}_{P}} & \equiv   \int \d \phi_\sT \cos 2\phi_\sT\,\frac{\d\sigma^{UP}}{\d y\,\d\xB\,\d^2 \bm{q}_{\sT} } =   \pi\, {\cal N}\, B^{UP}\, \frac{\bm q_\sT^2}{ M_p^2}\,h_1^{\perp\, g}(x, \bm q_\sT^2) \, .
\label{eq:D-Nb}
\end{align}
The hard scales of the process  $ e\,p \to e^\prime \,{\cal Q}\,X$ are the photon virtuality $Q$ and the quarkonium mass $M_{\cal Q}\approx 2 M_Q$. To avoid ratios of the two it is convenient to choose $Q=M_{\cal Q}$, but there are also advantages to varying $Q$. In the case of $P=U$ and writing $Q^2=cM_{\cal Q}^2$ we find that (as a function of the heavy-quark mass $M_Q$ rather than the quarkonium mass $M_\mathcal{Q}$)
\begin{align}
A^{UU} &= [1+(1-y)^2]{{\cal O}_{8}^S} + \left[2 (1-y)\, \frac{7+3c}{1+c} + y^2\, \frac{7+2c+3c^2}{(1+c)^2} \right] \frac{{\cal O}_{8}^P}{M_{Q}^{2}}  \,,\label{eq:AUU}\\
B^{UU} 
& = - (1-y)  \, \left [ {\cal O}_{8}^S - \frac{3-c}{1+c} \frac{{\cal O}_{8}^P}{M_Q^2} \right]\, ,
\label{eq:BUU}
\end{align}
where $ {\cal O}_{8}^S \equiv  \langle0\vert{\cal O}_{8}^{{\cal Q}}(^{1}S_{0})\vert0\rangle $  and $ {\cal O}_{8}^P \equiv  \langle0\vert{\cal O}_{8}^{{\cal Q}}(^{3}P_{0})\vert0\rangle $. These two long-distance matrix elements have been determined with quite large uncertainty from fits to data (see below), but in all cases ${\cal O}_{8}^P/M_Q^2 \leq {\cal O}_{8}^S$, in accordance with the scaling rules ${\cal O}_{8}^P \sim v^4$ and ${\cal O}_{8}^S\sim v^3$. Using the $Q^2$- and, to a lesser extent, $y$-dependence offers one method to extract the two matrix elements or at least their ratio if the gluon TMDs are not yet known. 
In $A^{UU}$, the prefactor multiplying ${\cal O}_{8}^P$ is largest when $y$ is smallest, and has a stronger $y$- and $c$-dependence than the one multiplying ${\cal O}_{8}^S$. Considering larger $Q^2$ will reduce the contribution from ${\cal O}_{8}^P$. More specifically, for $y=0.1$ and $c=1$ the prefactor multiplying ${\cal O}_{8}^P$ is approximately 5 times the one multiplying ${\cal O}_{8}^S$, whereas for large $c$ it reduces to a factor 3 independently of $y$. In $B^{UU}$,  the contribution from ${\cal O}_{8}^P$ is completely independent of $y$ and flips sign when $c=3$. Therefore, especially the $Q^2$-dependence could be exploited to change the relative contributions from ${\cal O}_{8}^S$ and ${\cal O}_{8}^P$ to the terms in Eqs.~\eqref{eq:D-Na} and~\eqref{eq:D-Nb}.

If one is able to determine the polarization of the quarkonium state, then there is another method to extract the two LDMEs. For longitudinally polarized quarkonium $P=L$ one obtains 
\begin{align}
A^{UL}
& = \frac{1}{3}\,[1+(1-y)^2] \, {\cal O}_{8}^S +  \left[2(1-y)\,\frac{1+10c+c^2}{(1+c)^2} +  y^2\,\frac{1+2c+c^2}{(1+c)^2} \right] \frac{{\cal O}_{8}^P}{M_{Q}^{2}}   \,, \label{eq:AUL}
\\
B^{UL}
& = (1-y) \left [ -\frac{1}{3}\, {\cal O}_{8}^S + \frac{{\cal O}_{8}^P}{M_Q^2} \right ]\, .
\label{eq:BUL}
\end{align}
Compared to $A^{UU}$, the ${\cal O}_{8}^P$ term in $A^{UL}$ has a different $y$ and $c$ dependence, which has quite important implications as we will discuss now. For $y=0.1$ and $c=1$ the prefactor multiplying ${\cal O}_{8}^P$ in $A^{UL}$ is around 9 times larger than the one multiplying ${\cal O}_{8}^S$, whereas for large $c$ it reduces to a factor 3 independently of $y$. This means that the prefactor of ${\cal O}_{8}^P$ in $A^{UL}$ can not only compensate for ${\cal O}_{8}^P/M_Q^2$ being smaller than ${\cal O}_{8}^S$ (${\cal O}_{8}^P$ could be smaller than ${\cal O}_{8}^S$ by as much as an order of magnitude), but  it can also lead to a significant deviation of $A_{UL}$ from $A_{UU}/3$ (and of $A_{UT}$ from $2A_{UU}/3$), signaling the production of polarized $J/\psi$ mesons. This would be in contradiction to the recent conclusion of Ref.~\cite{Qiu:2020xum},  i.e.\  that the dominance of the contribution of the  ${}^1S_0^{[8]}$ $c\bar{c}$ state to the cross section for the electron-hadron scattering process $e\,h \to J/\psi\, X$ implies that the produced $J/\psi$ meson will likely be unpolarized. This is presented as a robust test of NRQCD factorization and as a way to shed light on the $J/\psi$ production mechanism. As opposed to this statement from a collinear factorization analysis, our TMD factorization study indicates that the production of polarized $J/\psi$ mesons is not necessarily in contradiction with NRQCD, as it could simply signal a kinematic enhancement factor in front of the matrix element ${\cal O}_{8}^P$.

For $B^{UL}$ there is no $c$-dependence, only the overall $y$-dependence of $1-y$. Analogous expressions can be obtained when the $J/\psi$ meson is transversely polarized, namely
\begin{align}
A^{UT} &  = [1+(1-y)^2]  \left [\frac{2}{3} \, {\cal O}_{8}^S + \, 2\,\frac{3+ c^2}{(1+c)^2}\, \frac{{\cal O}_{8}^P}{M_Q^2} \right] \,, \label{eq:AUT}
\\
B^{UT} 
& = (1-y) \left [-\frac{2}{3}\,  {\cal O}_{8}^S +2\, \frac{1-c}{1+c}\, \frac{{\cal O}_{8}^P}{M_Q^2} \right] \, .
\label{eq:BUT}
\end{align}

Due to the fact that in most fits (see below) ${\cal O}_{8}^P/M_Q^2 \ll {\cal O}_{8}^S$, the $Q^2$-dependence in $A^{UP}$ and $B^{UP}$ is numerically strongly suppressed. Therefore, the experimental observation of $Q^2$-dependence in these observables would directly indicate the relevance of the ${\cal O}_{8}^P$ matrix element w.r.t. ${\cal O}_{8}^S$. Note that, through the prefactor in Eq.~\eqref{eq:N}, the cross sections fall of as $1/c^2$ at large $c$, where also experimental uncertainties will increase. Hence, the most relevant kinematic region for our purposes is the one where $c\approx 1$.

For $c=1$, we obtain for the production of unpolarized quarkonium states, from Eqs.~\eqref{eq:D-Na}, \eqref{eq:D-Nb},  \eqref{eq:AUU}, and \eqref{eq:BUU}: 
\begin{align}
D^{{\cal Q}_{U}}
& = \pi^3 \,   \frac{\alpha^2 \alpha_s e^2_Q}{ 8\, M_Q^5}  \,\left [ \frac{1+(1-y)^2}{y} \, {\cal O}_{8}^S+ \frac{10-10y+3y^2 }{y}\,
\frac{{\cal O}_{8}^P}{M_{Q}^{2}}  \right ] \,f_1^g(x, \bm q_\sT^2)\,, \\
N^{{\cal Q}_{U}} 
&= \pi ^3 \, \frac{\alpha^2 \alpha_s e^2_Q}{ 16\, M_Q^5} \left ( \frac{1-y}{y} \right  ) \, \left [-{\cal O}_8^S + \frac{{\cal O}_{8}^P}{M_Q^2} \right ]\,  \frac{\bm q_\sT^2}{ M_p^2}\,h_1^{\perp\, g}(x, \bm q_\sT^2)\,,  
\end{align}
while for longitudinally polarized quarkonium states, from Eqs.~\eqref{eq:D-Na}, \eqref{eq:D-Nb},  \eqref{eq:AUL}, and \eqref{eq:BUL}:
\begin{align}
D^{{\cal Q}_L}
& = \pi^3 \,   \frac{\alpha^2 \alpha_s e^2_Q}{ 8\, M_Q^5}  \,\left [ \frac{1}{3}\frac{1+(1-y)^2}{y} \, {\cal O}_{8}^S+ \frac{6-6y+y^2 }{y}\,
\frac{{\cal O}_{8}^P}{M_{Q}^{2}}  \right ] \,f_1^g(x, \bm q_\sT^2)\,, \\
N^{{\cal Q}_L} 
&= \pi ^3 \, \frac{\alpha^2 \alpha_s e^2_Q}{ 16\, M_Q^5} \left ( \frac{1-y}{y} \right  ) \, \left [-\frac{1}{3} {\cal O}_8^S + \frac{{\cal O}_{8}^P}{M_Q^2} \right ]\,  \frac{\bm q_\sT^2}{ M_p^2}\,h_1^{\perp\, g}(x, \bm q_\sT^2)\,. 
\end{align}
To our present LO accuracy in $\alpha_s$, the ratios $D^{{\cal Q}_L}/D^{{\cal Q}_{U}}$ and $N^{{\cal Q}_L}/N^{{\cal Q}_{U}}$ are independent of the TMDs:
\begin{align}
\frac{D^{{\cal Q}_L}}{D^{{\cal Q}_{U}}}
& = \frac{(1+(1-y)^2) \, {\cal O}_{8}^S/3+ (6-6y+y^2)\, {\cal O}_{8}^P/M_{Q}^{2}}{(1+(1-y)^2) \, {\cal O}_{8}^S+ (10-10y+3y^2)\,{\cal O}_{8}^P/M_{Q}^{2}}\,, \\
\frac{N^{{\cal Q}_L}}{N^{{\cal Q}_{U}}}
&= \frac{{\cal O}_8^S/3 - {\cal O}_{8}^P/M_Q^2}{{\cal O}_8^S - {\cal O}_{8}^P/M_Q^2}\,. 
\end{align}
These ratios offer another method to extract the two CO matrix elements if the polarization of the quarkonium state can be determined. 
Analogous ratios can be defined for transversely polarized quarkonium states,
\begin{align}
D^{{\cal Q}_T}
& = \pi^3 \,   \frac{\alpha^2 \alpha_s e^2_Q}{ 4\, M_Q^5}  \, \frac{1+(1-y)^2}{y}\left [ \frac{1}{3}\, {\cal O}_{8}^S+ 
\frac{{\cal O}_{8}^P}{M_{Q}^{2}} \right ] \,f_1^g(x, \bm q_\sT^2) \,, \\
N^{{\cal Q}_T} 
&= - \pi ^3 \, \frac{\alpha^2 \alpha_s e^2_Q}{ 24\, M_Q^5} \left ( \frac{1-y}{y} \right  ) \,  {\cal O}_8^S \, \frac{\bm q_\sT^2}{ M_p^2}\,h_1^{\perp\, g}(x, \bm q_\sT^2) \,,
\end{align}
see Eqs.~\eqref{eq:D-Na}, \eqref{eq:D-Nb},  \eqref{eq:AUT}, and \eqref{eq:BUT}. Notice that $N^{{\cal Q}_T}$ depends only on one of the two matrix elements, namely ${\cal O}_8^S$. Again, the ratios $D^{{\cal Q}_T}/D^{{\cal Q}_{U}}$ and $N^{{\cal Q}_T}/N^{{\cal Q}_{U}}$ are independent of the TMDs, as are the ratios $D^{{\cal Q}_T}/D^{{\cal Q}_{L}}$ and $N^{{\cal Q}_T}/N^{{\cal Q}_{L}}$.

The third and last method to be discussed is the one first suggested in Ref.~\cite{Bacchetta:2018ivt}. For this, we will compare the above expressions to the analogous ones for the process $e\, p \to e^\prime \,Q \,\overline{Q}\, X$, where $Q$ is either a charm or bottom quark. The differential cross section for the process 
\begin{equation}
 e(\ell) \,+\,p (P,S)\,\to \,e^\prime(\ell^\prime) \,+\, Q(K_Q)\,+\,\overline{Q}(K_{\overline Q})\,+\,X,
 \end{equation}
 in which the quark-antiquark pair is almost back to back in the plane orthogonal to the direction of the proton and the exchanged virtual photon, is written as~\cite{Boer:2010zf,Pisano:2013cya,Boer:2016fqd}
\begin{equation}
 \d\sigma^{Q\overline Q} \equiv \frac{\d\sigma^{Q \overline Q}}{\d z \,\d y\,\d\xB\,\d^2 \bm{K}_{\perp} \,\d^2 \bm{q}_{\sT} }\,.
\end{equation}
In the $\gamma^*p$ center-of-mass frame, the difference of the transverse momenta of the outgoing quark and  antiquark, $\bm K_\perp \equiv (\bm K_{Q\perp}-\bm K_{\overline Q \perp})/2$, should be much larger than the vector sum $\bm q_\sT \equiv \bm K_{Q\perp}+\bm K_{\overline Q \perp}$. The angles $\phi_\sT$ and $\phi_\perp$ are the azimuthal angles of $\bm q_\sT$ and $\bm K_\perp$ w.r.t.\ the lepton plane, respectively. Furthermore, $z = K_Q \cdot  P/ q\cdot P$. 

The process $e\, p \to e^\prime \,Q \,\overline{Q}\, X$ depends on three large scales: $M_Q$, $Q$ and $K_\perp \equiv \vert \bm K_\perp \vert$. 
Since $M_{{\cal Q}} \approx 2M_Q$, it is most convenient to choose $K_\perp = Q=2\, \sqrt{c}\,  M_Q$. With this choice we obtain
\begin{align}
D^{Q \overline Q} & \equiv  \int \d \phi_\sT \, \d \phi_\perp\,\frac{\d\sigma^{Q \overline Q}}{\d z \,\d y\,\d\xB\,\d^2 \bm{K}_{\perp} \,\d^2 \bm{q}_{\sT} }  = \pi \, \frac{\alpha^2 \alpha_s e^2_Q}{M_Q^4 \,c\,y\,z(1-z)} \, \frac{d^{Q\overline{Q}}}{2 [1+ 4 c(1+  z(1-z)) ]^3} \,f_1^g(x, \bm q_\sT^2)\,, \nonumber \\
N^{Q \overline Q} & \equiv
\int \d \phi_\sT\, \d \phi_\perp\cos 2\phi_\sT\,\frac{\d\sigma^{Q \overline Q}}{\d z\, \d y\,\d\xB\,\d^2 \bm{K}_{\perp}\d^2 \bm{q}_{\sT} }=   \pi \, \frac{\alpha^2 \alpha_s e^2_Q}{M_Q^4\,c\,y}  \, \frac{n^{Q\overline{Q}} }{2 [1+4c(1+ z(1-z))]^3} \,  \frac{\bm q_\sT^2}{ M_p^2}\,h_1^{\perp\, g}(x, \bm q_\sT^2)\,,
\label{eq:DNQQb}
\end{align}
where
\begin{align}
d^{Q\overline{Q}} & = \big(1+(1-y)^2\big) \Big[(1+4c)^2-2z(1-z)\big[1-4c\big(z^2+(1-z)^2\big)+8c^2 (2-9z+11z^2-4z^3+2z^4)\big]\Big]\\
&-128y^2c^2 z^2(1-z)^2\,, \nonumber\\
n^{Q\overline{Q}} & = - ( 1-y )[1+4 c\,z(1-z)]^2 \,.
\label{eq:d-nqqb}
\end{align}
In order to avoid the necessity for TMD evolution in the comparison to the quarkonium production case, we fix $c=1$, i.e.\ we take $K_\perp=Q= 2 M_Q$. Assuming furthermore that the energy is approximately equally distributed to the heavy quark and antiquark, we choose $z = 1/2$. Therefore, from Eqs.~\eqref{eq:DNQQb} and \eqref{eq:d-nqqb}, we get~\cite{Boer:2016fqd} 
\begin{align}
D^{Q \overline Q} & \equiv  \int \d \phi_\sT \, \d \phi_\perp\,\frac{\d\sigma^{Q \overline Q}}{\d z \,\d y\,\d\xB\,\d^2 \bm{K}_{\perp} \,\d^2 \bm{q}_{\sT} }  = \pi \, \frac{\alpha^2 \alpha_s e^2_Q}{54 \, M_Q^4} \,\left (  \frac{26 -26 y + 9y^2}{y} \right ) \,f_1^g(x, \bm q_\sT^2)\,,\\
N^{Q \overline Q} & \equiv
\int \d \phi_\sT\, \d \phi_\perp\cos 2\phi_\sT\,\frac{\d\sigma^{Q \overline Q}}{\d z\, \d y\,\d\xB\,\d^2 \bm{K}_{\perp}\d^2 \bm{q}_{\sT} }=   - \pi \, \frac{\alpha^2 \alpha_s e^2_Q}{108 \, M_Q^4}  \left ( \frac{1-y }{y} \right )  \frac{\bm q_\sT^2}{ M_p^2}\,h_1^{\perp\, g}(x, \bm q_\sT^2)\,. 
\end{align}

The comparison to quarkonium production then yields the following two independent observables
\begin{align}
{\cal R} & \equiv \frac{D^{{\cal Q}_{U}}}{D^{Q \overline Q}}= \frac{27\, \pi^2}{4}\,\frac{1}{M_Q} \, \frac{ [ 1+(1-y)^2 ] \, {\cal O}_{8}^S + ( 10-10y+3y^2  )\, {\cal O}_{8}^P /M_{Q}^{2}}{26 -26 y +9 y^2} \, , \\ 
{\cal R}^{\cos2\phi_\sT} & \equiv \frac{N^{{\cal Q}_{U}}}{N^{Q \overline Q}}= \frac{27 \pi^2}{4}\, \frac{1}{M_Q} \,  \left [{\cal O}_8^S - \frac{1}{M_Q^2} \, {\cal O}_8^P \right ] \,.
\label{eq:ratio-R}
\end{align}
Since they appear in two distinct combinations, the LDMEs ${\cal O}_8^S$ and ${\cal O}_8^P$ can be extracted from experimental measurements of the above ratios. It should be stressed that the presented expressions are LO  in $\alpha_s$ and order $v^3$ in the NRQCD velocity parameter $v$. Next-to-leading order (NLO) corrections in $\alpha_s$ will reintroduce sensitivity to the gluon as well as quark TMDs, which in turn will introduce a $q_\sT$-dependence in ${\cal R}$ and ${\cal R}^{\cos2\phi_\sT}$. Since the LO CO contribution, which we consider here, is expected to dominate over the NLO CS and CO contributions, parametrically by a factor $v^3 \pi/\alpha_s \approx 2$ and in practice by a larger factor for high $Q^2$~\cite{Fleming:1997fq}, we expect this dependence to be small. Another source of $q_\sT$-dependence in ${\cal R}$ and ${\cal R}^{\cos2\phi_\sT}$, however, may come from final state smearing effects. Indeed, in NRQCD the outgoing quarkonium state has the same transverse momentum as the $Q\overline{Q}$ pair, which is of course an idealization. While adding some additional uncertainty, the effects from smearing are expected to be mild \cite{Bacchetta:2018ivt}, entering merely as an overall $q_\sT$-dependent prefactor which is independent of the orbital angular momentum ($L$) of the CO  state, at least in the perturbative region \cite{Boer:2020bbd}. Thus, whenever a $q_\sT$-dependence is observed in ${\cal R}$ and/or ${\cal R}^{\cos2\phi_\sT}$, final state smearing effects in terms of shape functions~\cite{Echevarria:2019ynx,Fleming:2019pzj,Boer:2020bbd} should be considered next to the higher-order contributions from TMDs. Differences in the $q_\sT$-dependence of the ${\cal O}_8^S$ and ${\cal O}_8^P$ contributions would signal higher order terms or an $L$-dependence in the shape functions. 

In case the quarkonium polarization state can be determined, additional expressions can be obtained which can be used for consistency checks as they involve different combinations of the same CO matrix elements. For longitudinally polarized quarkonium production, one obtains: 
\begin{align}
{\cal R}_L & = \frac{9\, \pi^2}{4}\,\frac{1}{M_Q} \, \frac{ [ 1+(1-y)^2 ] \, {\cal O}_{8}^S + 3\, ( 6-6y+y^2  )\, {\cal O}_{8}^P /M_{Q}^{2}}{26 -26 y +9 y^2}  \,,\\
{\cal R}_L^{\cos2\phi_\sT} & =  \frac{27 \pi^2}{4}\, \frac{1}{M_Q} \,  \left [\frac{1}{3}\,{\cal O}_8^S - \frac{1}{M_Q^2} \, {\cal O}_8^P \right ] 
\,,
\end{align}
and for transversely polarized quarkonium production:
\begin{align}
{\cal R}_T & = \frac{9\, \pi^2}{2}\,\frac{1}{M_Q} \, \frac{ [ 1+(1-y)^2 ] \, {\cal O}_{8}^S + 3\, ( 2-2y+y^2  )\, {\cal O}_{8}^P /M_{Q}^{2}}{26 -26 y +9 y^2}  \,,\\
{\cal R}_T^{\cos2\phi_\sT}  & =   \frac{9\, \pi^2}{2}\, \frac{1}{M_Q} \,{\cal O}_8^S   \,.
\end{align}
These expressions satisfy:
\begin{align}
{\cal R}_L^{\cos2\phi_\sT} + {\cal R}_T^{\cos2\phi_\sT} &  = {\cal R}^{\cos2\phi_\sT} \,,\\
{\cal R}_L + {\cal R}_T &  = {\cal R} \,.
\end{align}
Note that the measurement of ${\cal R}_T^{\cos 2\phi_\sT}$  directly probes the matrix element ${\cal O}_{8}^S=  \langle0\vert{\cal O}_{8}^{{\cal Q}}(^{1}S_{0})\vert0\rangle $. Selecting photon virtualities such that $Q\simeq M_{\cal Q}$ ensures that this relation is not spoiled by large logarithmic corrections. The polarized quarkonium expressions also allow to assess the impact of higher order contributions, because these are different for the unpolarized and polarized cases and do not cancel in the ratios.  

In order to assess the potential of this third method we perform a numerical investigation of the various ratios in the next section.

\section{Numerical results}\label{sec:numerics}
In Tables \ref{tab:jpsiLDME} and \ref{tab:uLDME} we list various available determinations of the CO matrix elements.
\begin{table}[t]
\begin{centering}
\begin{tabular}{|c|c|c|c|c|c|c|}
\hline 
Fit No.\ & Reference & $\langle0|\mathcal{O}_{8}^{J/\psi}\bigl(^{1}S_{0}\bigr)|0\rangle$ & $\langle0|\mathcal{O}_{8}^{J/\psi}\bigl(^{3}P_{0}\bigr)|0\rangle/M_c^2$ & Units & $M_c$\tabularnewline
\hline 
\hline 
1 & Butensch\"on \& Kniehl~\cite{Butenschoen:2010rq}
& $4.50 \pm 0.72$ & $-0.54 \pm 0.16$ & $\times10^{-2}\,\mathrm{GeV}^{3}\mathrm{}$& 1.5 GeV\tabularnewline
2 & Chao {\it et al.}~\cite{Chao:2012iv} 
& $8.9 \pm 0.98$ & $0.56 \pm 0.21$ & $\times10^{-2}\,\mathrm{GeV}^{3}\mathrm{}$ & {\rm not specified} \tabularnewline
3 & Sharma \& Vitev~\cite{Sharma:2012dy}& $1.8 \pm 0.87 $ & $1.8 \pm 0.87$ & $\times10^{-2}\,\mathrm{GeV}^{3}$ & 1.4 GeV \tabularnewline
4 & Bodwin {\it et al.}~\cite{Bodwin:2014gia}
& $9.9 \pm 2.2$ & $  0.49 \pm 0.44 $ & $\times10^{-2}\,\mathrm{GeV}^{3}\mathrm{}$& 1.5 GeV\tabularnewline
\hline
\end{tabular}
\par\end{centering}
\caption{Fit values of the CO LDMEs from $J/\psi$ production (with some corrections w.r.t.\ \cite{Bacchetta:2018ivt}).}
\label{tab:jpsiLDME}
\vspace{0.4cm}
\begin{centering}
\begin{tabular}{|c|c|c|c|c|c|}
\hline 
Fit No.\ & Reference & $\langle0|\mathcal{O}_{8}^{\Upsilon(1S)} (^{1}S_{0}  )|0\rangle$ & $\langle0|\mathcal{O}_{8}^{\Upsilon (1S )}\bigl(^{3}P_{0}\bigr)|0\rangle/(5M_b^2)$ & Units & $M_b$ \tabularnewline
\hline  
\hline 
5 & Sharma \& Vitev~\cite{Sharma:2012dy}& $1.21 \pm 4.0$ & $1.21 \pm 4.0$ & $\times10^{-2}\,\mathrm{GeV}^{3}$& 4.88 GeV\tabularnewline
\hline 
\end{tabular}
\par\end{centering}
\caption{Fit values of the CO LDMEs from $\Upsilon(1S)$ production.}
\label{tab:uLDME}
\end{table}
\begin{table}[t]
\begin{centering}
\begin{tabular}{|c|c|c|c|c|c|c|}
\hline 
Fit No.\ & Reference & $~~~~~\sigma^\text{CO}~~~~~$ & $~~\sigma^\text{CO}/\sigma^\text{CS}~~$& $~~\sigma^\text{CO}/\sigma^\text{CS} \,(z>0.9)~~$ & $~~\sigma^\text{CO}/\sigma^\text{CS}\, (z>0.95)~~$  \tabularnewline
\hline 
\hline 
1 & Butensch\"on \& Kniehl~\cite{Butenschoen:2010rq}
& $\!\!\!-\!24.2$ pb& $\!\!\!\! -1.7$ & $\!\!\!\! -7.6$ & $\!\!\!\! -14$\tabularnewline
2 & Chao {\it et al.}~\cite{Chao:2012iv} 
& 27.2 pb & 2.2 & 9.7  & 18 \tabularnewline
3 & Sharma \& Vitev~\cite{Sharma:2012dy} & 83.3 pb  & 6.5 & 29 & 53 \tabularnewline
4 & Bodwin {\it et al.}~\cite{Bodwin:2014gia}
& 24.2 pb &   1.9  & 8.3& 15\tabularnewline
\hline
\end{tabular}
\par\end{centering}
\caption{The values for $\sigma^\text{CO}$ and $\sigma^\text{CO}/\sigma^\text{CS}$ for $J/\psi$ production, calculated at $y = 0.1, Q=2 M_c = 3$  GeV and $\xB = 0.01$, with different cuts on the variable $z$. The results are obtained using the fits for the CO LDMEs given in Table~\ref{tab:jpsiLDME}. For the CS matrix element, we adopted the values $1.32~ \text{GeV}^3$  (fit 1), $1.16~ \text{GeV}^3$ (fit 2), and $1.2~ \text{GeV}^3$ (fits $3,4$).}
\label{tab:CS-CO}
\vspace{0.4cm}
\begin{centering}
\begin{tabular}{|c|c|c|c|c|c|}
\hline 
Fit No.\ & Reference & $~~~~~\sigma^\text{CO}~~~~~$ & $~~\sigma^\text{CO}/\sigma^\text{CS}~~$& $~~\sigma^\text{CO}/\sigma^\text{CS} \,(z>0.9)~~$ & $~~\sigma^\text{CO}/\sigma^\text{CS}\, (z>0.95)~~$ 
 \tabularnewline
\hline  
\hline 
5 & Sharma \& Vitev~\cite{Sharma:2012dy}&~ $7.0\times 10^{-3}$ pb ~& 8.2 & 29 & 54 \tabularnewline
\hline 
\end{tabular}
\par\end{centering}
\caption{The values for $\sigma^\text{CO}$ and $\sigma^\text{CO}/\sigma^\text{CS}$ for $\Upsilon$ production, calculated at $y = 0.1, Q=2 M_b = 9.76$  GeV~\cite{Sharma:2012dy}, and $\xB = 0.09$, with different cuts on the variable $z$. The results are obtained using the fits for the CO LDMEs given Table~\ref{tab:uLDME}. For the CS matrix element, we adopted the value $10.9~\text{GeV}^3$.}
\label{tab:CS-CO-ups}
\end{table}
These have been used in the calculation of the cross section $\sigma^\text{CO}$ for the production of unpolarized $J/\psi$ and $\Upsilon$ mesons presented in Tables~\ref{tab:CS-CO} and \ref{tab:CS-CO-ups}, respectively. The results are obtained within the collinear factorization framework at LO in the strong coupling constant, in the kinematic region defined by  $y=0.1$, $Q= 2 M_Q$ and  $\sqrt{s}= 100$ GeV, which imply $\xB = 0.01$ for $J/\psi$ production ($M_c=1.5$ GeV) and $\xB = 0.09$ for $\Upsilon$ production ($M_b = 4.88$ GeV). Moreover, we have adopted the LO MSTW parametrization for the gluon distribution function~\cite{Martin:2009iq}, with the hard scale taken to be equal to $Q=2M_Q$. The ratio with the CS background, $\sigma^\text{CO}/\sigma^\text{CS}$, is shown as well. Note that all the LDME extractions were performed at NLO accuracy, except the Sharma \& Vitev set which is LO. The negative cross section corresponding to the NLO LDMEs of Butensch\"on \& Kniehl is due to the dominant contribution of the negative matrix element ${\cal O}_8^P$  for the process and the kinematic region under investigation. Although $\sigma^\text{CO}/\sigma^\text{CS}$ strongly depends  on the choice of the $J/\psi$ LDME set, we find that it is possible to suppress the CS contribution by requiring $z>0.9$. Such a cut will not affect the CO contribution, for which $z=1$ at LO and neglecting smearing effects. However, the results presented in Tables~\ref{tab:CS-CO} and \ref{tab:CS-CO-ups} should be regarded as indicative and taken with caution. Indeed, including smearing effects and/or NLO corrections to the hard part will soften the $z$-distribution of the CO contribution when $z \approx 1$, and lead to smaller ratios $\sigma^\text{CO}/\sigma^\text{CS}$ than the ones presented, especially when considering the $z>0.95$ cuts.

\begin{figure}[b] 
   \centering
   \includegraphics[width=2.2in]{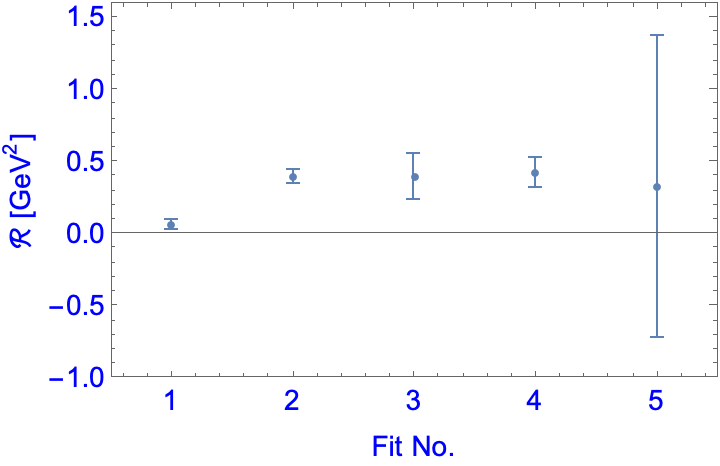} \
    \includegraphics[width=2.35in]{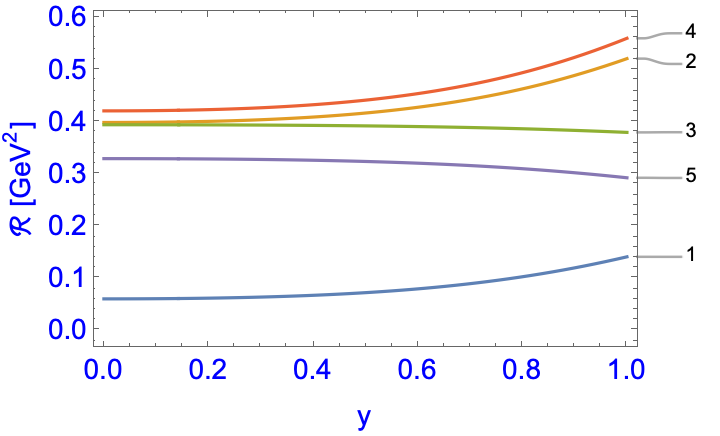} \
     \includegraphics[width=2.15in]{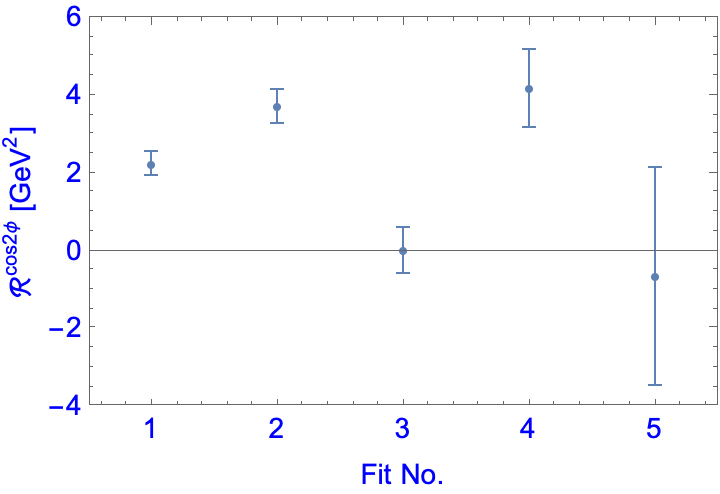} 
      \caption{The values for ${\cal R}$ and ${\cal R}^{\cos2\phi_\sT}$ obtained using the fits for the CO LDMEs given in Tables~\ref{tab:jpsiLDME} and~\ref{tab:uLDME}. ${\cal R}$ is shown for $y=0.1$ fixed in the left panel, and as function of $y$ in the middle one, for the central values of the 5 fits. The right figure shows ${\cal R}^{\cos2\phi_\sT}$ and holds for all $y$.}
      \label{fig:RU}
\end{figure}

The results for ${\cal R}$ and ${\cal R}^{\cos2\phi_\sT}$ are displayed in Fig.\ \ref{fig:RU}, where for fit 2 we assumed $M_c=1.5$ GeV. Possible correlations between the errors of the two fitted LDMEs have not been taken into account, i.e.\ the errors are simply added in quadrature. This may imply an overestimate of the uncertainty. 
From these results it is clear that different fits give quite different results. 
For instance, fit 1 gives a result compatible with zero for ${\cal R}$ but not for ${\cal R}^{\cos2\phi_\sT}$, whereas for fit 3 it is the other way around. These results give the general impression that ${\cal R}^{\cos2\phi_\sT}$ may be significantly larger than ${\cal R}$, perhaps even by as much as a factor of $10$.

Note that the ratios are not normalized to $[0,1]$ for ${\cal R}$ or to $[-1,1]$ for ${\cal R}^{\cos2\phi_\sT}$. However, one can relate their ratio to a ratio of asymmetries that {\it are} normalized:
\begin{align}
\frac{{\cal R}^{\cos 2 \phi_T}}{{\cal R}} = \frac{\langle \cos 2\phi_T \rangle_{{\cal Q}}}{\langle \cos 2\phi_T \rangle_{Q\overline{Q}}},
\end{align}
where $\langle \cos 2\phi_T \rangle_X = N^X/D^X$ for $X={\cal Q}$ or $X=Q\overline{Q}$.
From this one can see  that even though $\langle \cos 2\phi_T \rangle_{{\cal Q}}$ and $\langle \cos 2\phi_T \rangle_{Q\overline{Q}}$ are not known, a rough average of the fits shows that the $\cos 2\phi_T$ asymmetry could be substantially larger
in $J/\psi$ production than in open charm production. This would suggest that a study of the linearly polarized gluon TMD may be even more promising in $J/\psi$ production than in open charm production. Indeed, this seems to be supported by studies in a small-$x$ model, where $\langle \cos 2\phi_T \rangle_{Q\overline{Q}}$ was found to be at the $5-10\%$ level \cite{Boer:2016fqd} and $\langle \cos 2\phi_T \rangle_{{\cal Q}}$ around $10-20\%$ \cite{Bacchetta:2018ivt}, in both cases increasing with increasing values of $q_\sT$. Since we consider gluon induced processes, smaller $x$ values may be beneficial because the gluon TMDs would be enhanced. On the other hand, nonlinear QCD evolution was seen to have a suppressing effect on the $\langle \cos 2\phi_T \rangle_{{\cal Q}}$ asymmetry as $x$ becomes smaller \cite{Bacchetta:2018ivt}. At the EIC, the smaller the $x$ value, the smaller the $Q$ values covered, and the smaller the $q_\sT$-range for which TMD factorization is expected to hold. Therefore, one has to keep a balance between the $x$- and $Q$-ranges.

\begin{figure}[t] 
   \centering
   \includegraphics[width=2.2in]{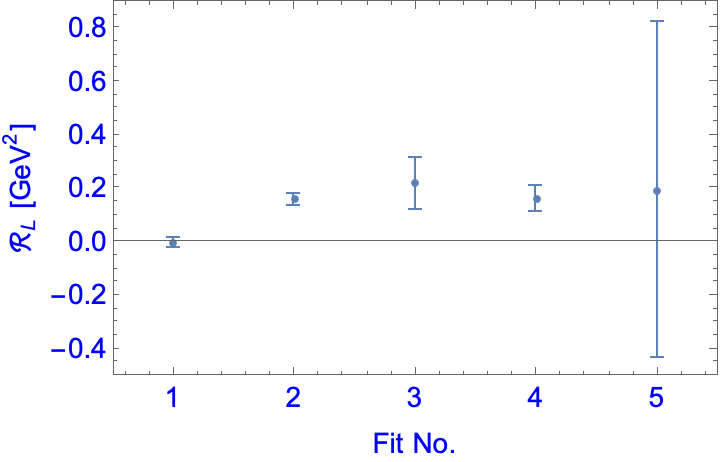} \
    \includegraphics[width=2.35in]{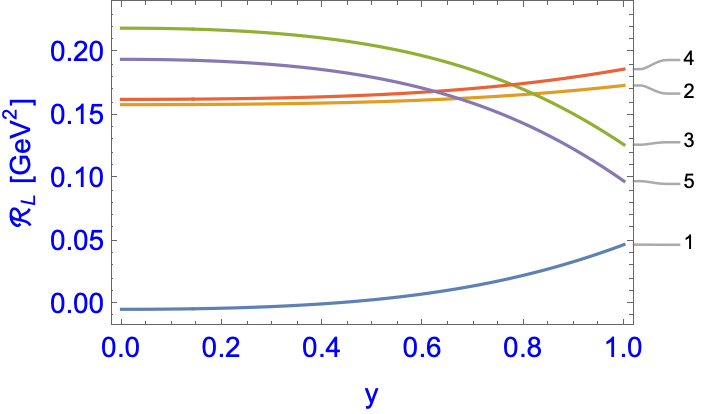} \
     \includegraphics[width=2.15in]{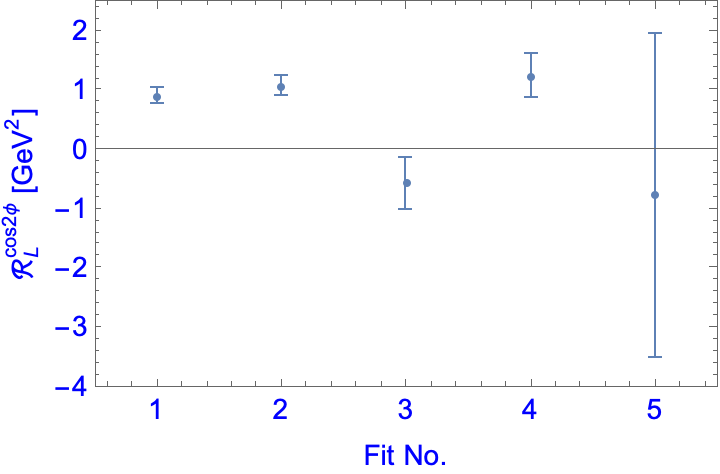} 
      \caption{The values for ${\cal R}_L$ and ${\cal R}_L^{\cos2\phi_\sT}$ obtained using the fits for the CO LDMEs given in Tables~\ref{tab:jpsiLDME} and~\ref{tab:uLDME}. ${\cal R}_L$ for $y=0.1$ fixed is shown in the left panel, and as function of $y$ in the middle one, for the central values of the 5 fits. The right figure shows ${\cal R}_L^{\cos2\phi_\sT}$ and holds for all $y$.}
      \label{fig:RL}
\end{figure}
\begin{figure}[b] 
   \centering
   \includegraphics[width=2.2in]{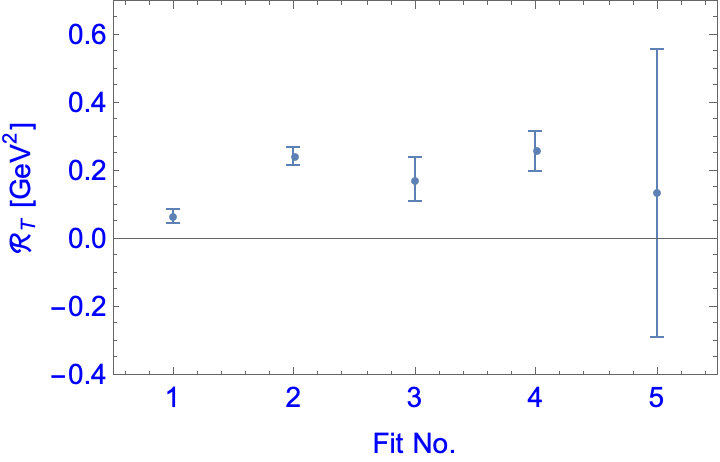} \
    \includegraphics[width=2.35in]{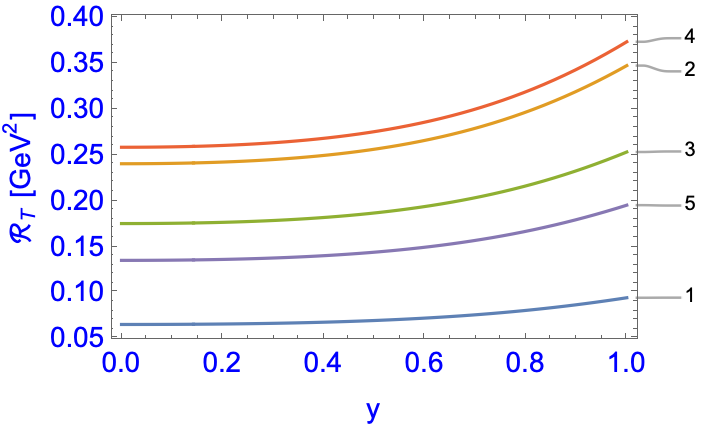} \
     \includegraphics[width=2.15in]{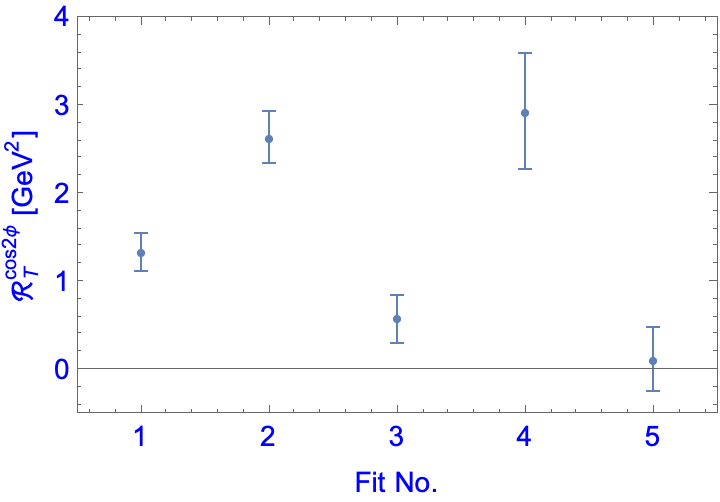} 
      \caption{The values for ${\cal R}_T$ and ${\cal R}_T^{\cos2\phi_\sT}$ obtained using the fits for the CO LDMEs given in Tables~\ref{tab:jpsiLDME} and~\ref{tab:uLDME}.  The left figure shows ${\cal R}_T$ for $y=0.1$, and the middle figure as function of $y$ for the central values of the 5 fits. The right figure shows ${\cal R}_T^{\cos2\phi_\sT}$ and holds for all $y$.}
      \label{fig:RT}
\end{figure}
Similarly, the results for the $P=L$ and $P=T$ cases are given in Figs.\ \ref{fig:RL} and \ref{fig:RT}.
The figures show only a moderate $y$-dependence, most pronounced for large $y$-values.  
Note that although ${\cal R}_P^{\cos2\phi_\sT}$ has no $y$-dependence, both numerator and denominator have a prefactor $(1-y)$, hence vanish at $y=1$. It is also worth remarking that, for the $\Upsilon$ meson, the error bars for ${\cal R}_T^{\cos2\phi_\sT}$ are significantly smaller than those for ${\cal R}_L^{\cos2\phi_\sT}$ and ${\cal R}^{\cos2\phi_\sT}$. This is due to the fact that in the $P=T$ case, this ratio only depends on the LDME ${\cal O}_8^S$.

We end this section with a comment on the $Q^2$ or $c$-dependence of the various ${\cal R}$ ratios. The cross sections in both the numerator and the denominator fall off as $1/c^2$ for large $c$, therefore, this dominant behavior cancels. Nevertheless, the ratios exhibit a stronger $c$-dependence than the moderate dependence of $A^{UP}$ and $B^{UP}$. This leads to larger differences between different fits for increasing $c$. However, as mentioned, at larger $c$ the experimental uncertainty will also increase, therefore, for the ${\cal R}$ ratios there is only limited advantage in exploiting the $c$-dependence over a broad range above $c=1$. Considering different $c$-values can nevertheless put further constraints on the fits.

\section{Summary and Conclusions \label{sec:conc}}
In this paper we have presented several methods to use $J/\psi$ and $\Upsilon$ production in unpolarized SIDIS at the EIC to obtain improved determinations of two CO LDMEs, $\langle0\vert{\cal O}_{8}^{J/\psi}(^{1}S_{0})\vert0\rangle$ and $\langle0\vert{\cal O}_{8}^{J/\psi}(^{3}P_{0})\vert0 \rangle$, which are currently poorly known. The first method exploits the $y$- and $Q^2$-dependence of the $\phi_\sT$-integrated and $\cos 2\phi_\sT$-weighted $\phi_\sT$-integrated cross sections. The second method exploits the polarization states of the produced quarkonium state, and is independent of the gluon TMDs. The third method is based on the comparison of the process $ e\,p \to e^\prime \,J/\psi \,(\Upsilon)\,X$ with $e\, p \to e^\prime \,D \,\overline{D}\, (B \,\overline{B})\, X$. Estimates based on the available fits of the CO matrix elements were presented for this third method. Since the available fit values are not compatible with each other, any new method to determine the CO matrix elements with better precision will be worth exploring. Despite the large uncertainty the estimates show sizable ratios. These results suggest that the $\cos 2\phi_T$ asymmetry that arises from linearly polarized gluons inside the unpolarized proton could be substantially larger in $J/\psi$ production than in open charm production. For typical EIC kinematics, this is corroborated by the small-$x$ model studies in Refs.~\cite{Boer:2016fqd} and \cite{Bacchetta:2018ivt}, where asymmetries around 5-10\%  and 10-20\% were found for open charm and $J/\psi$ production, respectively.

The experimental determination of the polarization states of the produced quarkonia will not only be very helpful in the determination of the CO matrix elements, it could also help improve the understanding of the quarkonium production mechanism in NRQCD, which still poses challenges~\cite{Brambilla:2010cs,Andronic:2015wma,Lansberg:2019adr}. In contrast to the recent conclusion of Ref.~\cite{Qiu:2020xum} that $J/\psi$ mesons produced in electron-hadron collisions will likely be unpolarized and thus can provide a rigorous test of NRQCD, we find,  from our LO TMD factorization analysis, that polarized production of $J/\psi$ mesons is not necessarily a stringent NRQCD prediction.

The robustness of our results is ensured by the following considerations. First, TMD factorization is applicable to SIDIS in the kinematic region where $q_\sT^2\ll Q^2$. Considering an outgoing quarkonium state instead of a light meson is not expected to spoil the factorization, provided one allows for a $q_\sT^2$-dependence of the LDMEs (the shape functions, Refs~\cite{Echevarria:2019ynx,Fleming:2019pzj,Boer:2020bbd}).
However, in Ref.~\cite{Boer:2020bbd} it was shown that for large $q_\sT^2$, the final state smearing is independent of the orbital angular momentum of the bound state, hence the same for the $S$- and $P$- wave CO states, and thus drops out in the ratios at LO. We have assumed this to hold for all $q_\sT^2\ll Q^2$, but otherwise the restriction $M_p^2 \ll q_\sT^2\ll Q^2$ could be considered.
Second, based on the velocity scaling rules, NLO CS contributions are suppressed by a factor $v^3 \pi/\alpha_s \approx 2$ w.r.t.\ the LO CO contributions which we consider here. In an actual numerical study (in the collinear framework), we showed that this suppression can be even stronger, especially for the LDME sets of Ref.~\cite{Sharma:2012dy}, and the CS contributions can be made negligible if one imposes a lower cut on $z$. In turn, the possible diffractive background that becomes important at high $z$, as well as potential corrections from higher-twist effects,  can be suppressed by looking at sufficiently high virtualities $Q^2$~\cite{Fleming:1997fq}. Third, the asymmetries that we propose are always the ratios of two cross sections. Hence, besides
being independent of the normalizations of the cross sections, they are also expected to be less sensitive to higher-order corrections or to other sources of uncertainties such as the exact value of the charm and bottom mass or the final state smearing effects.  
The importance of higher order corrections and final state smearing effects will make themselves apparent in the possible $q_\sT$-dependence of the various ratios that were presented, and can be included as needed.

We conclude that our findings show that heavy-quark final states at EIC are very promising tools to provide improved determinations of CO LDMEs.

\acknowledgments
This project has received funding from the European Union's Horizon 2020 research and innovation programme under grant agreement STRONG 2020 - No 824093.
C.P.\  also acknowledges financial support by Fondazione di Sardegna under the project {\em Proton tomography at the LHC}, project number F72F20000220007 (University of Cagliari).

\end{document}